\documentclass[preprint,superscriptaddress,preprintnumbers,amsmath,amssymb,pre]{revtex4}
\usepackage{epsfig}
\usepackage{latexsym}
\usepackage{dcolumn}
\usepackage{color}
\usepackage{subfigure}
\usepackage{bm}
\begin{document}
\date{\today}
\title{Snow metamorphism: a fractal approach}

\author{Anna Carbone}\email{anna.carbone@polito.it}
 \affiliation{Physics Department and CNISM, Politecnico di Torino,\\ Corso Duca degli
Abruzzi 24, I-10129 Torino, Italy}

\author{Bernardino M. Chiaia}\email{bernardino.chiaia@polito.it}
\affiliation{Department of Structural and Geotechnical Engineering, Politecnico di Torino,\\  Corso Duca degli Abruzzi 24, I-10129 Torino, Italy}

\author{Barbara Frigo}\email{barbara.frigo@polito.it}
\affiliation{Department of Structural and Geotechnical Engineering, Politecnico di Torino,\\  Corso Duca degli Abruzzi 24, I-10129 Torino, Italy}

\author{Christian T\"{u}rk}\email{christian.turk@polito.it}
\affiliation{Physics Department and CNISM, Politecnico di Torino,\\ Corso Duca degli Abruzzi 24, I-10129 Torino, Italy}

\begin{abstract}
Snow is a porous disordered  medium consisting of air and three water phases: ice, vapour and liquid. The ice phase consists of an assemblage of grains, ice matrix, initially arranged over a random load bearing skeleton. The quantitative relationship between density and morphological characteristics of different snow microstructures is still an open issue. In this work, a three-dimensional fractal description of density corresponding to different snow microstructure is put forward.  First, snow density is simulated in terms of a generalized Menger sponge model. Then, a fully three-dimensional compact stochastic fractal model is adopted. The latter approach yields a quantitative map of the randomness of the snow texture, which is described as a three-dimensional fractional Brownian field with the Hurst exponent $H$  varying as continuous parameters. The Hurst exponent is found to be strongly dependent on snow morphology and density. The approach might be applied to all those cases where the morphological evolution  of snow cover or ice sheets should be conveniently described at a quantitative level.
\\
\end{abstract}

\maketitle
\section{Introduction}
Physical and mechanical properties of snow greatly vary over space and rapidly evolve in time (snow metamorphism), affecting thermal conductivity, strength, heat capacity, density and texture \cite{Dash,Blackford,Colbeck}.
Density and texture depend on meteorological conditions, temperature, pressure and  humidity, which, together with the mechanical strains cause metamorphism
and,  ultimately, instability of snow \cite{Dadic,Golubev,Frolov,Kaempfer1,Kaempfer2,Lowe,Vetter,Nakamura,Flin,Heggli,Arakawa}.
Density is the parameter used for classifying and assessing snow properties, thanks to the simplicity of in-situ measurements and can be adopted as a parameter for quantifying characteristics  as viscosity, shear stress and strength, cohesion and mechanical properties, such as Young's Modulus and Poisson's ratio.
Usually, researchers refer to the specific density, defined as the ratio of the snow density $\rho_{snow}$ to the ice density $\rho_{ice}$=917 kg/m$^3$ and ranging between 0.05 to 0.60, to describe mechanical properties of snow.  However, different types of snow having the same density might exhibit
completely different mechanical features.
Elucidating the quantitative relation between density and mechanical characteristic of snow is an open issue worthy of investigation beyond
purely speculative interests. For example, the knowledge of density of added or lost snow is needed in altimeter measurements of dynamic thinning 
of Arctic and Antarctic ice sheets and monitoring of incipient avalanches \cite{Pritchard,Gay,Floyer}.
\par
Scaling properties and fractional calculus have been extensively adopted to characterize different classes of
materials \cite{Turcotte}. In particular, fractal concepts have been already used for modelling snow crystals and reproducing snowflake morphology (see e.g. the von Koch snowflake curves and the
 application of general iterated function systems). In the last two decades, fractal theories of snow have resulted in several applications: from the measure of fractal dimension of images, to remote sensing and mapping of snow cover and depth distribution by satellite and Lidar images, from the study of the roughness of the snowpack, to the determination of the air flux across snow surface and to the definition of
the spatial variation of the snow water equivalent  \cite{Deems1,Deems2,Emerson,Mehrota,Fassnacht1,Fassnacht2,Rees,Manes}.  The fractal character of snow has been recently studied by the Centre Etudes de Neige of M\'et\'eo France  \cite{www,Faillettaz} by analysing three-dimensional tomography obtained for cubic samples with $2.5\,\mathrm{mm}$ size and different densities.  Relationships between
mechanical properties (tension, shear strength, toughness) and specific density in terms of power laws with  non integer exponents  suggesting  fractal features of snow are reported in \cite{Kirchner,Miller,Petrovic}.
Furthermore, fractal theory has been  used  to investigate the stability of the snow cover at larger scales. Based on the renormalisation group model, the probability of occurrence of snow avalanche events, assuming scale invariance of the snowpack at the smallest
scales with a consequent implication of the same behavior at the largest scales \cite{Chiaia1,Chiaia2}.
\par
The present work is addressed to investigate the multiscale character of snow density, by adopting a fractal description of the distribution of ice grains, able to reproduce the local randomness of real microstructure.
The paper is organized as follows.
In Section II, snow density is simulated by means of a generalized Menger sponge model, characterized by a discrete set of Hurst exponent values.
In Section III, a fully three-dimensional fractional Brownian model is reported. This model has the advantage to reproduce the randomness of the local
microstructure of snow samples in a more realistic way. Snow density is mapped to the three-dimensional fractional Brownian field,
with the Hurst exponent $H$ continuously ranging from 0 to 1.
The proposed model provides a fully three-dimensional analysis in terms of a continuum fractional Brownian field rather than the discrete Menger model description  \cite{Carbone1,Turk}. In the framework of this model, fractal dimension $D$ and  Hurst exponent $H$ quantify ice distribution and reproduce the values of snow density for different microstructures.
This fully three-dimensional fractal model can be used to relate snow texture, obtained by in-situ measurements o remote imaging techniques,
and density changes, important for monitoring snow cover and ice sheets dynamics, through the estimate of Hurst exponent at different snow metamorphism stages.
\section{Menger sponge model}
\begin{figure}[htbp]
\centering
\subfigure[\label{a}]%
{\includegraphics[width=3cm,angle=0]{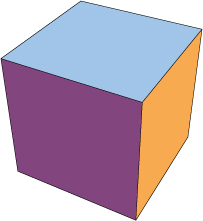}}
\subfigure[\label{a}]%
{\includegraphics[width=3cm,angle=0]{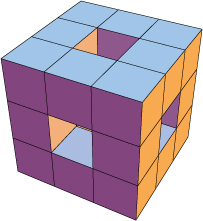}}
\subfigure[\label{b}]%
{\includegraphics[width=3cm,angle=0]{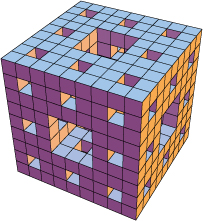}}
\caption{\label{Figure1} Solid homogeneous cube at the initial stage of the iteration process (a), Menger sponge at the first iteration (b),  Menger sponge at the second iteration (c). Size, density and porosity are respectively:  $r_o$ , $\rho_o$ and  $\phi_o = 0$ (a);  $r_1 =  3r_o$, $\rho_1 = (20/27) \rho_o$ and  $\phi_1 = 7/27$ (b);  $r_2 = 9 r_o$, $\rho_2 = (400/729) \rho_o$ and  $\phi_2 = 329/729$ (c) \cite{Turcotte}.}
\end{figure}

In this Section, the fractal, known as Menger sponge, is used to characterize scale invariant features of porosity and density of snow.
The Menger sponge is generated as shown in the scheme of Fig.1. First, the bulky cube (a) is divided into $3 \times 3 \times 3 = 27$ equal subcubes (b). Then, 7 subcubes are removed from the center of each face and from the center of the cube, resulting in $N_f=20$ filled subcubes and $N_e = 3^3 - N_f=7$ empty subcubes. This single step is repeatedly applied to the remaining cubes.
For a solid cube with linear size $r_0$,  the first step Menger sponge is characterized by linear size $r_1 = 3r_0$ (Fig.~\ref{Figure1}(b)).
 The second step Menger sponge is characterized by linear size $r_2 = 9r_0$  as shown in Fig.~\ref{Figure1}(c).
 In general,  $r_i = (3^i) r_0$ is the linear dimension of the fractal cube (Menger sponge)  at the iteration $i$.
 \par
 The fractal dimension of the Menger sponge is given by  $D=\ln{N_f}/\ln{3}\simeq2.727$. Here,
we assume self-similarity which implies a linear relationship $H = 3 - D \simeq 0.273$ between fractal dimension $D$ and Hurst coefficient $H$.
 \par
The Menger sponge has been widely used to model porous media, whose relevant parameter is the porosity $\phi$ \cite{Turcotte}.  The porosity  $\phi_i$, defined as the relative volume of voids per unit volume, can be expressed by the following relationship:

\begin{equation}
\label{1}
\phi_i=1-\left(\frac{\rho_i}{\rho_0}\right) \hspace{8pt} ,
\end{equation}
where $\rho_0$ is the initial density of the bulky cube and $\rho_i$ is the density of the Menger sponge at the iteration $i$.
By taking into account that  $\rho_i$ and $\rho_0$  are inversely proportional to the volumes and, then, to the linear sizes $r_0$ and $r_i$, Eq.~(\ref{1}) can be written as:
\begin{equation}\label{2}
\phi_i=1-\left(\frac{r_0}{r_i}\right)^{3-D} \hspace{8pt} ,
\end{equation}
where $D$ is the fractal dimension, that is the scale-independent parameter characterizing the morphology of a porous material.
Therefore, density and void index of the
Menger sponge, as a function of the linear dimensions of the solid cube $r_0$ and fractal cube $r_i$\, can be written at the $i^{th}$ iteration as:
\begin{equation}\label{3a}
\frac{\rho_i}{\rho_0}=\left(\frac{r_0}{r_i}\right)^{3-D}=\left(\frac{r_0}{r_i}\right)^{3-\frac{\ln{N_f}}{\ln{3}}}=\left(\frac{r_0}{r_i}\right)^{H} \hspace{5pt}.
\end{equation}					
For the Menger sponge shown in Fig.~\ref{Figure1}(b),  the porosity  is $\phi_1 = 7/27$ and the density is $\rho_1 =  20 \rho_0/27$, while
for the  Menger sponge of Fig.~\ref{Figure1}(c),  the porosity is $\phi_2 = 329/729$ and the density is $\rho_2 =   400 \rho_0/729$.
\par
The above described procedure can be generalized by removing an arbitrary number $N_e$ of subcubes (instead of 7) out of an arbitrary number of solid cubes $N_f$ (instead of 20). This generalized construction
results in fractal structures with Hurst exponent different than $H=0.273$.
\par To model snow samples,  an homogeneous ice cube characterized by density $\rho_{0} = \rho_{ice} = 917 \,\,\mathrm{kg/m}^3$ and linear size $r_{0} = r_{grain}$ is considered at the initial step. Then, snow  is obtained as a Menger sponge, i.e. a
 fractal form of ice,  characterized by density $\rho_{snow} = \rho_i$, and linear size $r_{snow}= r_{i}$ at the iteration $i$.
 Thus, Eq.~(\ref{3a}) is rewritten for snow as follows:
\begin{equation}\label{4}
\frac{\rho_{snow}}{\rho_{grain}}=\left(\frac{r_{grain}}{r_{snow}}\right)^{3-\frac{\ln{N_f}}{\ln{3}}}=\left(\frac{r_{grain}}{r_{snow}}\right)^{3-D}=\left(\frac{r_{grain}}{r_{snow}}\right)^{H} \hspace{5pt}.
\end{equation}				
\par
Fractal dimension $D$ and Hurst exponent $H$  can be taken as a measure of porosity of the snow sample.
The fractal character of snow can be analysed by using images obtained on various cubic samples with same size and different densities, showing the granular structure and the spatial distribution of its voids. By means of the box-counting method, the fractal dimension of four snow samples, characterized by grains with different diameter, could be determined \cite{Faillettaz}.
By applying the generalized Menger sponge model the fractal dimension $D$ of different classes of snow can be calculated (Table I). These values have been obtained by using the same size of the samples:  $r_{snow} = r_i = 100\,\mathrm{mm}$  and the grain size $r_{grain}$ ranging between $0.05\,\mathrm{mm}$ and $0.25\,\mathrm{mm}$.  By analyzing  density and void index as a function of the linear sample dimension $r_i$ (grain size), we observe that as the grain size increases, snow differs more and more from ice. We also observe that the values of the fractal dimension $D$ measured by the box-counting method, ranging between  $D=2.62$ to  $D=3$ \cite{Faillettaz}, are consistent with the values calculated by the  Menger sponge model (Table I).
At small scales, ice and snow approximately show the same behavior while the spatial variability of the density does not greatly influence the mechanical properties. Therefore, we argue that snow density is a function of the scale and the probability to find large defects (e.g. super-weak zones in a weak layer) increases with the dimension of the snow grain, as for example providing more intrinsic brittleness for large snow slopes \cite{Chiaia1,Chiaia2}. Numerical results reported in Table I confirm that $D$ is an accurate measure of the distribution of the ice mass into snow samples.

\section{Three-dimensional Fractional Brownian model}
 In the previous section, the generalized Menger sponge model, characterized by scale invariant porosity, is used to develop a fractal description of snow density of samples with different microstructures. This description is of significant practical relevance, however it does not fully capture the fractal structure of snow samples. One drawback is due to the discreteness of fractal dimension. Moreover, the porosity of a real-world fractal should be free of scale requiring an infinite number of iterations for generating the sponge. On account of these limitations, in order to describe snow as a sintered porous material consisting of a continuous ice network, a generalization of the random midpoint displacement algorithm is implemented to model snow as a three-dimensional fractal heterogeneous medium \cite{Turk}.
The method for generating compact fractal disordered media relies on fractional Brownian functions, which are characterized by correlation  depending on the distance $r$ as a power law. The approach is based on the function:
\begin{equation}
\label{rf1}
f_{H,j}(r) =
\frac{1}{2^d}\sum_k f_k(r) + \sigma_{j,d},
\end{equation}
with $r = (i_1,i_2,i_3)$.   The sum is calculated over the $k$ endpoints of the lattice and the quantity $\sigma_{j,d}$ is a random variable
defined at each iteration $j$ as:
\begin{equation}
\label{rf2}
\sigma^2_{j,d} = \sigma_0^2\left(\frac{\sqrt{d}N}{2^j}\right)^{2H}\left[1-2^{2(H-d)}\right],
\end{equation}
where the quantity $\sigma_0$ is drawn from a Gaussian distribution with zero mean and unitary variance. The Hurst exponent $H$, ranging from 0 to 1, is the input of the algorithm which is implemented at each iteration $j$ according to the following procedure. Initially, the lattice
is fully homogeneous, with the function $f_H(r)$ describing the fractal property taken as a constant, e.g. $f_H(r) = 0$. Then,
the values of the function $f_{H,j}(r)$ are seeded as random variables at the eight vertices of the cube.
The value assigned to the central point is obtained by means of Eqs.~(\ref{rf1},\ref{rf2}), by using the eight vertices as input.
The value at the center of each face is assigned in the same way, but with the sum  calculated over the  four vertices corresponding to each face.
Finally, the midpoint values of each of the twelve edges are calculated with the sum calculated over the vertices at the end points of the edges.
The first iteration of this algorithm results in 27 subcubes. These steps, except the initial seeds of the eight vertices, are iteratively repeated
for each of the 27 subcubes. Eventually, the number of subcubes will become $(3^j)^d$, where $j$ is the iteration number and $d=3$. Further details about this construction can be found in \cite{Turk}.
\par
\begin{figure}
\includegraphics[width=14cm,angle=0]{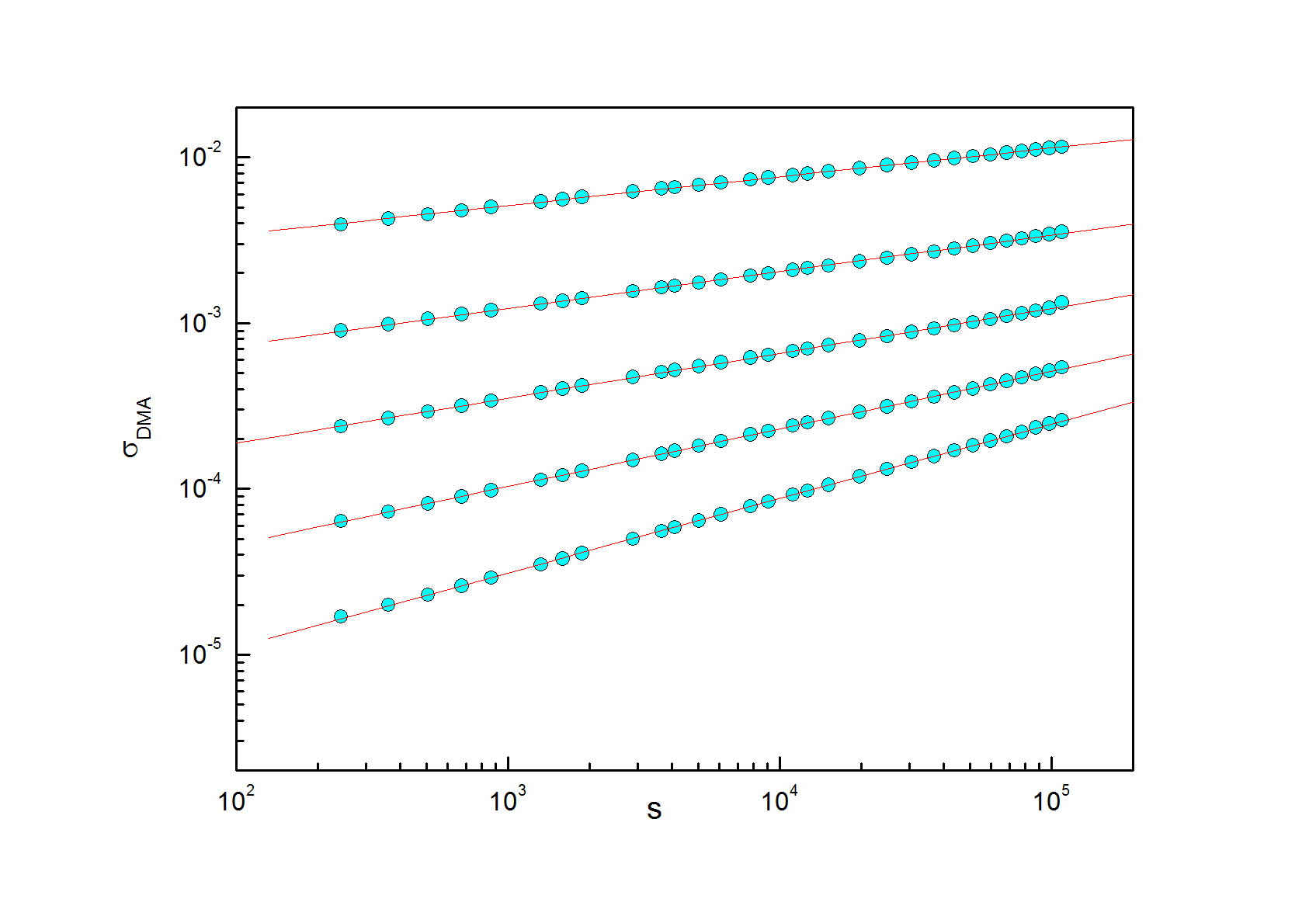}
\caption{\label{figure2}  Log-log plot of the function $\sigma_{DMA}$ as a function of scale $s$. The Hurst exponent is estimated by the slope of the best fit (red lines), respectively $H=0.1$, $H=0.2$, $H=0.3$, $H=0.4$ and  $H=0.5$. }
\end{figure}

By using the detrending moving average (DMA) algorithm \cite{Carbone1}, the Hurst exponent of the  fractal
structure can be subsequently estimated. The core of the DMA algorithm is the generalized variance $\sigma^2_{DMA}(s)$, that for $d=3$ writes:
\begin{equation}\label{DMAd}
\sigma^2_{DMA}(s) = \frac{1}{V}\sum_V\left[ f_H(r) - \tilde{f}_{n_1,n_2,n_3}(r)\right]^2,
\end{equation}
where $f_H(r) = f_H(i_1,i_2,i_3)$ is the fractional Brownian field with $i_1=1,2,...,\,N_1$, $i_2=1,2,...,\,N_2$ and $i_3=1,2,...,\,N_3$. The function $\tilde{f}_{n_1,n_2,n_3}(r)$ is given by
\begin{equation}\label{MAd}
\tilde{f}_{n_1,n_2,n_3}(r) = \frac{1}{\nu} \sum_{k_1} \sum_{k_2} \sum_{k_3} f_H(i_1-k_1,i_2-k_2,i_3-k_3),
\end{equation}
with the size of the subcubes $(n_1,n_2,n_3)$ ranging from $(3,3,3)$ to the maximum values $(n_{1max},n_{2max},n_{3max})$.  $\nu = n_1n_2n_3$ is the volume of the subcubes. The quantity
$V = (N_1-n_{1max})(N_2-n_{2max})(N_3-n_{3max})$ is the volume of the fractal cube over the average $\tilde{f}$ is defined. Eqs.~(\ref{DMAd}) and (\ref{MAd}) are defined for any geometry of the subarrays. In practice, it is computationally more suitable to use $n_1=n_2=n_3$ to avoid spurious directionality and biases in the calculations. In Fig.~\ref{figure2}, the log-log plots of $\sigma^2_{DMA}(s)$ vs $s$ are shown for fractal cubes generated according to the above described procedure. The cubes have Hurst exponent  $H=0.1$, $H=0.2$, $H=0.3$, $H=0.4$ and  $H=0.5$ respectively. The log-log plots of
$\sigma^2_{DMA}(s)$ as a function of $s$ are straight lines according to the power-law behavior:
\begin{equation}
\sigma^2_{DMA}(s) \propto (n_1^2 + n_2^2 + n_3^2)^H \propto s^H,
\end{equation}
because of the fundamental property of fractional Brownian functions.
\par
By using these algorithms, the fractality of the snow structure can be related to the snow density by mapping the fractional Brownian field $f_H(r)$ to a density field $\rho(r)$. In this framework, the Hurst exponent, varying as a continuous parameter, should be  intended as an index of specific snow compactness. Different snow textures have been simulated by varying the minimum value of the density $\rho_{min}$ between $0$ and $917 \,\mathrm{kg/m}^3$, while the maximum density is constant and equal to the ice density $\rho_{max}=\rho_{ice}=917$ kg/m$^3$.
\par
\begin{figure}[htbp]
\centering
\subfigure[\label{b}]%
{\includegraphics[width=5cm,angle=0]{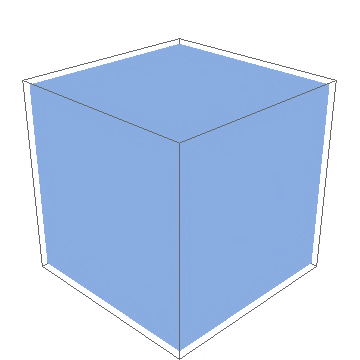}}
\subfigure[\label{a}]%
{\includegraphics[width=5cm,angle=0]{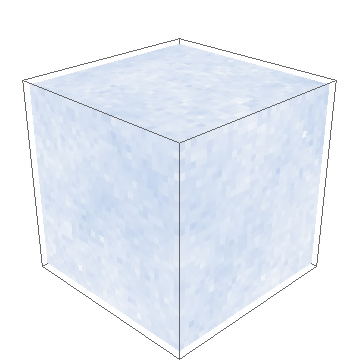}}
\caption{ \label{figure3} Fractal cubes generated according to the three-dimensional fractional Brownian model presented in Section III. The Hurst exponent is $H=0.1$ and the density $\rho(r)$ ranges respectively between $800-917\,\, \mathrm{kg/m}^3$ (a) and  $0-917\,\,\mathrm{kg/m}^3$ (b). The ratio between the cube edge and the grain size is 400, implying that there are $400 \times 400 \times 400$  values in each cube.}
\end{figure}
\begin{figure}
\includegraphics[width=14cm,angle=0]{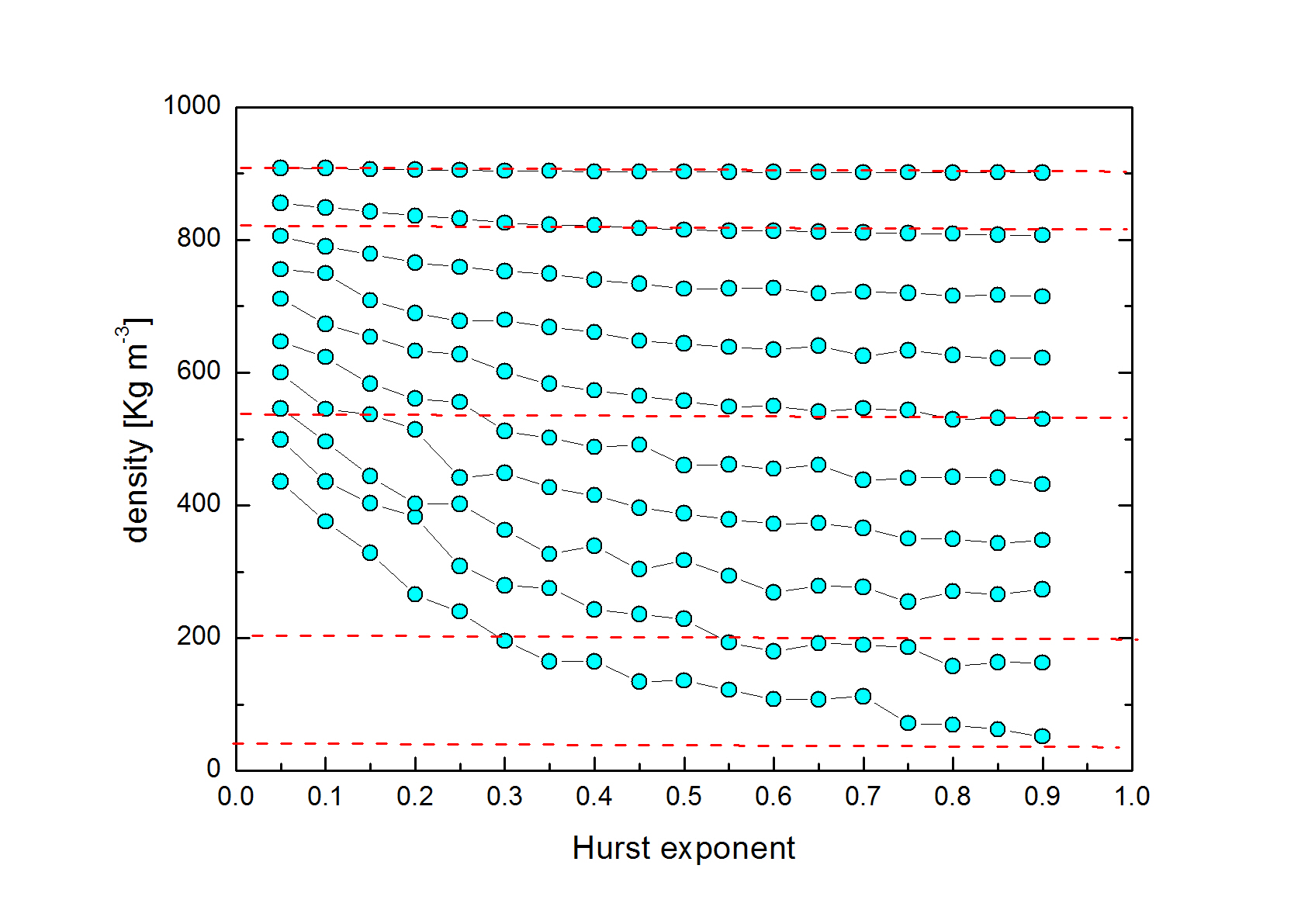}
\caption{\label{figure4}  Snow density as a function of the Hurst exponent. The different curves have been obtained by generating fractional Brownian functions defined over a cubic lattice \cite{Turk}. Then, the fractional function has been mapped to a density function. In order to simulate different snow microstructures, the minimum value of the density $\rho_{min}$ has been varied between $0$ and $917$ kg/m$^3$, while the maximum value of the density is kept constant and equal to the ice density  $\rho_{max}=\rho_{ice}=917$ kg/m$^3$. Different curves, from bottom to top, correspond to values of  $\rho_{min}$ ranging from  $0$ to $917$ kg/m$^3$, with step $100$ kg/m$^3$. Dashed horizontal lines indicate density ranges for dry snow ($50 < \rho < 200$); snow ($200 < \rho< 550$); firn ($550 <  \rho < 820$) and porous ice ($820 <  \rho < 917$) after \cite{Colbeck}.  }
\end{figure}

In Fig.~\ref{figure3} snow structures corresponding to cubes with size $r_0 = 100\, \mathrm{mm}$ and granular size $r_{grain} =  0.25\,\mathrm{mm}$,
with $H = 0.1$  are shown. The density ranges from $900$ to $917 \,\,\mathrm{kg/m}^3$ (a) and  from $0$ to $917 \,\, \mathrm{kg/m}^3$(b).
One can observe the difference between the more compact solid ice structure (a) and the almost fully porous media featured by several areas of lower density (b). Colors are scaled in such a way that darker areas correspond to higher densities.
Finally, the average density $\rho_{average}$ of the fractals generated according to the above procedure has been calculated. The results are plotted in Fig.~\ref{figure4}.
One can notice that the average density decreases more rapidly with lower values of $\rho_{min}$ as the Hurst exponent increases,
while the average density is practically unchanged as $H$ is changed, by taking $\rho_{min}$  close to the value $917 \, \mathrm{kg/m}^3$.
The present approach might have interesting applications for monitoring ice losses and snow metamorphism. By independent measures of snow density and
Hurst exponent, one can map the morphological evolution of snow/ice by using curves similar to those of Fig.~\ref{figure4}.
\section{Conclusions}
Physical and mechanical properties of snow are usually defined in terms of the specific density thanks to its simplicity of in-situ measurements. Unfortunately, the density is not univocally related to the snow microstructure, since different snow microstructure might exhibit the same global density.
 A scale invariant parameter is needed to quantify snow metamorphism in terms of the multiscale properties of snow density and porosity. We have proposed a fractal model for snow density based on  (i) a generalized Menger sponge and (ii) a stochastic fractional Brownian field. The present approach shows that different Hurst exponents correspond to the same value of density, implying that density alone does not yield complete information about snow microstructure.
 Nonetheless, thanks to this model, one should be able to investigate how the local structure evolves according to the fractal dimension in relation to other physical properties. The present work is the first step towards the investigation of the scaling properties of snow in a fully three-dimensional fractal framework, relevant to the validation of experimental results,
such as those reported in \cite{Gay,Nakamura}, and the description of the physical and mechanical properties which are of great interest for many application areas.
\section{acknowledgement}
We acknowledge financial support by: Fondazione CRT Progetto Alfieri; the Operational Programme Italy-France (Alps-ALCOTRA) Project ``DynAval-Dynamique des Avalanches: d\'epart et interactions \'ecoulement/obstacles"; Regione Piemonte;  Regione Autonoma Valle d'Aosta and Politecnico di Torino. Furthermore, CINECA is gratefully acknowledged for CPU time at the High Performance Computing (HPC) environment.

\clearpage

\clearpage

\clearpage

\clearpage

\clearpage

\clearpage
\begin{table}
\caption{\label{tab:1} Classification of snow and ice (first column) according to the density (second column).  The values of the Hurst exponent $H$ have been calculated for a sample with linear size $r_{snow} = 100\, \mathrm{mm}$  for different grain sizes $r_{grain}$ on the basis of the generalized Menger sponge model discussed in Section II. The snow type and density are taken from the classification reported in  \cite{Colbeck}.  }
\begin{ruledtabular}
\begin{tabular}{ccccc}
Snow type  & Density [$\mathrm{kg/m^3}$]  & $r_{grain}=0.05\,\mathrm{mm}$       & $r_{grain}=0.15 \,\mathrm{mm}$    & $r_{grain}=0.25\,\mathrm{mm}$ \\
\hline
\hline
Dry snow&	$50 < \rho < 200$ &	$0.3827< H < 0.2003$	  &	$0.4474 < H < 0.2342$  &		$0.4855 < H < 0.2542$\\  \hline
Snow&	$200 < \rho< 550$	& $0.2003 < H< 0.0673$	  &	$0.2342 < H < 0.0786$  &		$0.2542 < H < 0.0853$\\  \hline
Firn&	$550 <  \rho < 820$	& $0.0673 < H < 0.0147$  &	$0.0786 < H < 0.0172$  &		$0.0853 < H < 0.0187$ \\  \hline
Porous ice&	$820 <  \rho < 917$ &	$0.0147 < H < 0.0000$	  &	$0.0172< H < 0.0000$  &		$0.0187 < H < 0.0000$\\
\end{tabular}
\end{ruledtabular}
\end{table}


\begin{thebibliography}{<50>}


\bibitem{Dash}
J.G. Dash, A.W. Rempel and J.S. Wettlaufer, Rev. Mod. Phys. \textbf{78}, 695 (2006).

\bibitem{Blackford}
J.R. Blackford, J. Phys. D Appl. Phys. \textbf{40}, R355 (2007).

\bibitem{Colbeck}
S.C. Colbeck, J. Appl. Phys. \textbf{ 84}, 4585 (1998); S. Colbeck, E. Akitaya,  R. Armstrong, H. Gubler,  J. Lafeuille, K. Lied,
D. McClung,  and E. Morris,  The International classification for seasonal snow on the ground,  37  (1985).

\bibitem{Dadic}
R. Dadic, M. Schneebeli,   M. Lehning,  M. A. Hutterli,  and A. Ohmura, J. Geophys. Res. \textbf{113},  D14303 (2008).



\bibitem{Frolov}	
A.D. Frolov, and I.V. Fedyukin,   Annals of Glaciology \textbf{26}, 55 (1998).

\bibitem{Golubev}	
V.N. Golubev,  and A.D. Frolov, Annals of Glaciology \textbf{26},  45 (1998).


\bibitem{Kaempfer1} T.U. Kaempfer, M.  Schneebeli,   and   S.A. Sokratov,  Geophys. Res. Lett. \textbf{32},  L21503 (2005);
T.U.  Kaempfer, M. A. Hopkins  and  D.K. Perovich,  J. Geophys. Res. \textbf{112}, D24113  (2007).

\bibitem{Kaempfer2}
T.U. Kaempfer,  and M. Schneebeli,   J. Geophys. Res. \textbf{112},  D24101 (2007);
T.U. Kaempfer and M. Plapp,  Phys. Rev. E \textbf{79},  031502 (2009).

\bibitem{Lowe}  H. L\"{o}we, L. Egli, S. Bartlett, M. Guala, and C. Manes,  Geophys. Res. Lett., \textbf{34}, L21507,  (2007).


\bibitem{Vetter}
R. Vetter, S. Sigg, H.M. Singer, D. Kadau, H.J. Herrmann, M. Schneebeli, EPL  \textbf{ 89}, 26001  (2010).


\bibitem{Nakamura}	
T. Nakamura, O. Abe, T. Hasegawa, R. Tamura,  and Ohta, T. Cold Regions Science and Technology, \textbf{32},  13 (2001).

\bibitem{Flin}
F. Flin, J.B. Brzoska,  B. Lesaffre,  C. Col\'eou, R.A. Pieritz,   J. Phys. D: Appl. Phys. \textbf{36}, A49 (2003).

\bibitem{Heggli}
 M. Heggli, E. Frei  and  M. Schneebeli,
J. of Glaciology, \textbf{55},  631 (2009).
\bibitem{Arakawa}
H. Arakawa, K. Izumi, K. Kawashima and T. Kawamura
Cold Regions Science and Technology \textbf{59} 163 (2009).



\bibitem{Pritchard}
H.D. Pritchard, R.J. Arthern, D.G. Vaughan, L.A. Edwards,
Nature \textbf{461}, 971 (2009).

\bibitem{Gay}	
M. Gay,  M. Fily, C. Genthon,  M. Frezzotti, H. Oerter, J.G. Winther,  J. of Glaciology, \textbf{48}, 527 (2002).

\bibitem{Floyer}
J. Floyer and B. Jamieson, Cold Reg. Sci. Technol.  \textbf{59},  185 (2009).

\bibitem{Turcotte}
D.L. Turcotte,  Cambridge University Press, 20 (1997).

\bibitem{Deems1}
J.S. Deems, S.R. Fassnacht,  and  K.J. Elder, J. of Hydrometeor. \textbf{7}, 285 (2006).

\bibitem{Deems2}
J.S. Deems, S.R. Fassnacht, and  K.J. Elder, J. of Hydrometeor. \textbf{9},  977 (2008).
%
\bibitem{Emerson}
C.W.  Emerson, N. S. Lam,  and D.A. Quattrochi,  Photogrammetric Engineering \& Remote Sensing  \textbf{65},  51 (1999).

\bibitem{Mehrota}
R. Mehrotra, D. Kumar, Phys. Rev. Lett. \textbf{92},  254502 (2004).

\bibitem{Fassnacht1}
S.R. Fassnacht  and  J.S. Deems, Hydrol. Process. \textbf{20}, 829 (2006).

\bibitem{Fassnacht2}
S.R. Fassnacht, M.W. Williams and M.V. Corrao, Ecological Complexity \textbf{6},  221 (2009).

\bibitem{Rees}
W.G. Rees, N.S. Arnold, J. of Glaciology \textbf{52}, 214 (2006).

\bibitem{Manes}
C. Manes, M. Guala, H. L\"{o}we, S. Bartlett, L. Egli, and M. Lehning,
Water Resour. Res., \textbf{44}, W11407,  (2008).

\bibitem{www}
\textcolor[rgb]{0.00,0.00,1.00}{www.cnrm.meteo.fr/passion/neige1.htm}

\bibitem{Faillettaz}
 J. Faillettaz, F. Louchet, J.R. Grasso, Phys. Rev. Lett. \textbf{93},  208001   (2004); Faillettaz J. Le déclenchement des avalanches de plaque de neige: de l'approche mécanique à l'approche statistique. PhD thesis, Université Joseph Fourier (Grenoble I), 2003.

\bibitem{Kirchner}
H. O. K. Kirchner, H. Peterlik and G. Michot, Phys. Rev. E \textbf{69}, 011306 (2004);  J. Schweizer, G. Michot, H.O.K. Kirchner, Annals of Glaciology  \textbf{38}, 1 (2004);
H.O.K. Kirchner,  G. Michot,  H. Narita, and T. Suzuki, Phil. Mag. A \textbf{81} (9),  2161 (2001).
%
\bibitem{Miller}
D.A. Miller, E.E. Adams, J. of Glaciology \textbf{55}, 1003 (2009).
%
	
\bibitem{Petrovic}
J.J. Petrovic,   J. Mat. Sc. \textbf{38},  1 (2003).



\bibitem{Chiaia1}	
B. Chiaia, P. Cornetti and B. Frigo, Cold Regions Science and Technology \textbf{53}, 170  (2008).

\bibitem{Chiaia2}
B. Chiaia,  and B. Frigo, J. of Stat. Mech.: Theory and Experiment, P02056  (2009).




\bibitem{Carbone1}	
A. Carbone, Phys. Rev. E \textbf{76},   056703 (2007).

\bibitem{Turk}
C. T\"{u}rk,  A. Carbone, and  B.M. Chiaia,  Phys. Rev. E \textbf{81}, 026706 (2010).



\end{thebibliography}
\end{document}